\documentclass[a4paper,UKenglish,cleveref, autoref, thm-restate]{oasics-v2021}
\usepackage{todonotes}

\graphicspath{{./images/}}

\title{The HaMSE Ontology: Using Semantic Technologies to support Music Representation Interoperability and Musicological Analysis}

\titlerunning{The HaMSE Ontology} 

\author{Andrea Poltronieri\footnote{Corresponding author.}}{Department of Modern Languages, Literatures and Cultures (LILEC), University of Bologna, Italy}{andrea.poltronieri@studio.unibo.it}{https://orcid.org/0000-0003-3848-7574}{}

\author{Aldo Gangemi}{Department of Classical Philology and Italian Studies (FICLIT), University of Bologna, Italy}{aldo.gangemi@unibo.it}{https://orcid.org/0000-0001-5568-2684}{}

\authorrunning{A. Poltronieri and A. Gangemi} 

\Copyright{Andrea Poltronieri and Aldo Gangemi} 



\category{} 

\relatedversion{} 

\nolinenumbers 

\EventEditors{}
\EventNoEds{2}
\EventLongTitle{First International Workshop on Multisensory Data and Knowledge at the 3rd Conference on Language, Data and Knowledge (2021)}
\EventShortTitle{MDK 2021}
\EventAcronym{MDK}
\EventYear{2021}
\EventDate{September 1--3, 2021}
\EventLocation{Zaragoza, Spain}
\EventLogo{}
\SeriesVolume{}
\ArticleNo{5}
\funding{This project has received funding from the European Union's Horizon 2020 research and innovation programme under grant agreement No 101004746.}

\begin{document}

\maketitle

\begin{abstract}
The use of Semantic Technologies - in particular the Semantic Web - has revealed to be a great tool for describing the cultural heritage domain and artistic practices.
However, the panorama of ontologies for musicological applications seems to be limited and restricted to specific applications. 

\noindent{In this research, we propose HaMSE, an ontology capable of describing musical features that can assist musicological research. More specifically, HaMSE proposes to address issues that have been affecting musicological research for decades: the representation of music and the relationship between quantitative and qualitative data. To do this, HaMSE allows the alignment between different music representation systems and describes a set of musicological features that can allow the music analysis at different granularity levels.} 
\end{abstract}

\section{Introduction}
\label{sec:introduction}

Since the second half of the 20\textsuperscript{th} century, musicological research, along with a myriad of other disciplines, has been affected by a massive transition to computational technologies. Many aspects of the research have since begun to be entrusted to computers.

However, this transition seems to have affected only some sub-domains of musicological study. Even today, in fact, entire fields of musicology rely very little on computational tools to support research. This is partly due to the lack of means that can relate elements typically processed by computers, i.e. quantitative data, with information typically sought by research fields more related to the humanities field. 

In this research, current problems related to the musicological domain are analysed, together with the state-of-the-art technologies in computational musicology. In the end, an approach based on Semantic Technologies - in particular the Semantic Web - is proposed to improve and enhance musicological research.

A set of musical features that can serve this purpose is discussed, divided into four main categories: melodic, harmonic, structural and emotional. Different Music Information Retrieval approaches aimed at extracting such features are examined and used as requirements for the design of the ontology. Finally, the HaMSE ontology is proposed as a model for representing musical content in different forms (both symbolic and signal representations), linking them to the set of previously defined features.

\section{Challenges in Musicological Research}
\label{sec:challenges}

Music is a vivid field of research that embodies several subjects of interest and, at the same time, has always been a hard subject to analyse.

Marc Leman refers to music as a ``transdisciplinary'' field. Transdisciplinary, perhaps even more than the term ``interdisciplinary'', suggests that music cannot be fully understood by a single discipline, or by different disciplines that are just put next to each other without much interaction \cite{leman2008}. Musicology is the scholarly discipline that aims at analysing all these distinct elements, trying to encompass them all.

For the purposes of this research, both aspects of research methodology in musicology and the methods of representation used in music are analysed. Particular attention is given to the computational applications of both aspects, i.e. the fields in which computational tools are used for musicological analysis and computational representation of music.

\subsection{Methodological Challenges}

Musicology literally means ``the study of music''. In 1955, the American Musicological Society described it as a field of knowledge having as its object the investigation of the art of music as a physical, psychological, aesthetic, and cultural phenomenon. However, in the last two decades an important shift has occurred, that is, from music as an art (or art object) to music as a process in which the performer, the listener, and music as sound play a central role \cite{honing2004}. 

One of the most important internal categorizations of musicology has been made by Guido Adler in 1884, in the writing published as the opening article of the first journal of musicology.
In the article, the Austrian musicologist defined the internal subdivision of musicology into historical and systematic branches \cite{mugglestone1981}. 

The same subdivision of the discipline has been maintained by the more recent – even contemporary, scholars and researchers, with the addition of one more field: comparative musicology. During the middle of the twentieth century, this term was supplanted by ``ethnomusicology'', reflecting a new belief that cultural practices could only be understood concerning the particular societies that gave rise to them \cite{clarkecook2005}. 

One of the aims of this research is to analyse and highlight the main methodological differences between the research in systematic musicology and in the fields more related to the historical branch, and how they relate to each other. 

Systematic musicology is an umbrella term for subdisciplines of musicology that are primarily concerned with music in general, rather than specific manifestations of music. It has traditionally been conceived as an interdisciplinary science, whose aim is to explore the foundations of music from different points of view, such as acoustics, physiology, psychology, anthropology, music theory, sociology, and aesthetic \cite{leman1997}. Although systematic musicology was established as an academic discipline only in the 20\textsuperscript{th} century, the origins of the mathematical and scientific investigation in music can be dated back to the Greek antiquity. Parncutt \cite{parncutt2007} observes that until about 1600 musical thought was almost all ``systematic'', and it could be classified into theoretical, mathematical, philosophical, aesthetic, acoustical, psychological or sociological branches.

Nowadays, the dichotomy between systematic musicological research and historical musicological research implies an important methodological diversity.
Systematic musicology – in particular defined in its scientific facet called \emph{Naturwissenschaften} - is primarily empirical and data-oriented: it involves empirical psychology and sociology, acoustics, physiology, neurosciences, cognitive sciences, and computing and technology. On the other side, Historical musicology is typically included in the humanities domain. 

This dualism is attributable to the inner bivalence of the music material. Music is primarily constituted by sound, and hence its physical nature comes to our attention. The final result, though, is a human’s artistic creation, studied by humanistic disciplines as such. This involves an opposition – which sometimes leads to a contradiction, both in the methodology and in the epistemology of the two disciplines, which makes the analysis of music even more difficult.

From an epistemological point of view, the difference between humanities and science involves the tension between subjectivity and objectivity in all scholarship: while the humanities tend to address the creative products of the human spirit, the sciences tend instead to address the physical and biological environment of human beings - their natural surroundings \cite{parncutt2007}. Another difference between the two fields of study can be sought in the objective of the research. While scientists tend to seek general conclusions that can theoretically be as impersonal as possible, humanists tend to strive for complex and specific findings, that can possibly enrich the subject of their research.

However, the most seminal epistemic distinction between the sciences and the humanities (that somehow affects all the taxonomies described so far) is the one between data-rich and data-poor areas of research. Data-rich disciplines are in principle able to uncover or assemble as much information, evidence, observations, etc. as they wish, limited only by financial resources. Data-poor disciplines, on the other side, have little control over the volume of pertinent data \cite{huron1999}. 

In the second half of the 20\textsuperscript{th} century, musicological research began to make use of IT tools. This transition, however, took several decades to be realised, and can still be considered ongoing. Furthermore, it stands out the differentiation concerning ``music information science'' and ``musicology''. It seems that the two fields developed independently and, until at least 2005, on two separates tracks. The computational research about music was computer scientists’ jurisdiction, while musicology was on its own analogical steps. 

We assisted, then, to an expropriation of the musicological research put in place by computer science and engineering departments. This was reflected by the fact that most of the researchers in these fields have little (or even any) musical knowledge. A major reason that can explain this fact lies in an internal lack of systematization in musicology. Such a lack can find an explanation both in the fragmentation of the field in plenty of subfields and in the absence of a solid formalized theory. This was also the main objection moved by the sceptics about computational musicology, especially during the 80s. 

Conversely, historical musicology was (and still is) a subfield more resistant to technological contamination. Historical musicologists claim an underlying methodological incompatibility with these new computational approaches. Modern ``critical'' approaches have generally claimed that musical creations are first and foremost to be explained from their contexts. The new emphasis on subjectivity and originality in research, however, was achieved at the expense of evidence-based, incremental and collaborative research approaches \cite{volk2011}.

The question that arises is whether these two apparently antithetical approaches can find a way of communicating and mutually take advantage from each other. More specifically, the question is whether the study of music in the humanities - and therefore, a data-poor field - can benefit from a data-rich approach. The other question is if, vice versa, research fields more focused on the use of quantitative data can benefit from humanistic studies.

\subsection{Challenges in Music Representation}

One of the greatest challenges in musicology is related to the music representation. The complexity of this issue certainly reflects the multifaceted and twofold nature of music itself depicted in the section above. 

The main challenge, again, is to hold together the empirically measurable elements of music with the non-empirical ones. Moreover, a further layer of complexity is due to the manifold sources we can deal with when studying music, i.e. printed notation, raw sound, etc. In addition, the performing nature of music implies that different executions of the same piece may generate structurally different music data. 

Historically, the innovation that allowed the study of music and hence the rise of musicology has been the introduction of music notation. Music scores were initially introduced for the main purpose of recording music by giving other musicians the possibility of playing the same piece in turn, actually reproducing it. However, notation brought with it a wide range of issues, many of which are still unresolved. Music scores, in fact, symbolize rather than represent music \cite{cook2006}. The problems that afflict music representation concern numerous aspects of musical content. For example, a music representation system must take into account the multiple hierarchies that are created within the musical content (e.g. between voices, sections, phrases, chords), as well as the mapping between real time (measured in seconds) and metric time (measured in beats). Furthermore, there are several musical aspects that are still extremely difficult to represent, starting with timbre \cite{dannenberg1993}.  

Hugues Vinet \cite{vinet2004} proposes to categorise music representation systems into four categories, namely symbolic representation, signal representation, physical representation and knowledge representation. Physical representations result from physical descriptions of music phenomena, they include all the special characteristics of sound and are based on mathematical formalism, as well as signal representations. 

Symbolic representations describe music by means of entities that have an explicit musical meaning and, encoded in some digital format, can be parsed by a computer \cite{muller2015}.

On the other side, knowledge representation provides a structured formalization of useful knowledge on musical object for specific applications, such as music multimedia libraries. The resulting taxonomy can be synthesized as follows, in a scale that goes from concrete to abstract descriptions: 
\begin{romanenumerate}
\item Physical representations
\item Signal representations
\item Symbolic representations
\item Knowledge representations
\end{romanenumerate}
A representation system can be evaluated by means of a system based on two orthogonal dimensions. These two dimensions have been named as \emph{expressive completeness} and \emph{structural generality}. Expressive completeness is described as the extent to which the original content of the represented music may be retrieved and/or recreated from the representation knowledge \cite{wiggins2009}. Structural generality, instead, refers to the range of high-level structures that can be represented and manipulated \cite{wiggins1993}.

Among all the others, the most complete representation, i.e. the one with the biggest information quantity, corresponds to analog signal. Even though raw sound waveform allows the maximal expressive completeness, at the same time it provides us with almost no structural generality. Hence, for extracting structured information out of a soundwave we need to process that signal.

On the other hand, symbolic representations may allow a higher level of structural generality, but never convey the same degree of expressiveness of signal representations. A requirement for representing music symbolically is the creation of a system that defines how to represent every musical aspect consistently. A symbolic music representation system (MSR) may be suited for many different purposes, and its usefulness is relative to the task at hand. 

Different types of symbolic representation have been analysed for this research, divided into two main categories: standard symbolic representations and graph representations. The first category includes the most commonly used MSRs. These appear as a large and extremely heterogeneous group, in which MSRs designed for different needs, developed using different technologies, and describing different musical aspects can be found. 

MIDI, for example, is a data communication protocol that describes a means for music systems for exchanging information and control signals \cite{rothstein1995}. If we evaluate MIDI files considering expressive completeness and structural generality, it will score badly in both parameters. Considering expressive completeness, even though time is represented in “real” terms (i.e. milliseconds) and would allow for a good expressiveness, the pitch information is inevitably abstracted. This is due to the approximation of pitches to the piano keys. 

The ABC Notation \cite{walshaw2021} and the Chord Notation \cite{harte2005} are text-based music notation system, which aims to describe the music at the surface level of the single note and chords, respectively. 

**kern \cite{huron2002}, instead, is an MSR explicitly developed for musicological analysis as part of the Humdrum toolkit. LilyPond \cite{nienhuys2003}, on the other side, is a modular, extensible and programmable compiler for producing high-quality music notation: its main purpose is to bring the aesthetics of traditionally engraved music to computer printouts. 

The CHARM system \cite{harris1991} is presented as a general MSR, where the musical surface of this system was chosen to be at the level of notes and phrases. Another explicit choice in this MSR was to represent music in performance rather than music as scored. 

Finally, MusicXML \cite{good2001} is an XML-based music interchange language. It is intended to support the interchange between musical notation, performance, analysis, and retrieval applications. Instead, the Music Encoding Initiative (MEI) \cite{roland2002} is a community-driven, open-source effort to define a system for encoding musical documents in a machine-readable structure, recording physical and intellectual characteristics of music notation documents expressed as an eXtensible Markup Language (XML) schema. While both MusicXML and MEI encode music notation in XML, the latter also encodes information about the notation and its intellectual content in a structured and systematic way, retaining the structure and semantics of the notation.   

On the other hand, graph representations arise from the assumption that music is multiply hierarchical \cite{lerdahl1996}. The multiply hierarchical nature of music data implies that a tree-structure of the systems presented so far is not enough for representing such a level of hierarchical dependencies. In other words, defining music as multiply hierarchical means that the structure is, formally, a graph. In this category we can include the Spectra Graph \cite{pinto2007}, in which melodies are represented as graphs, based on the intervals between the notes they are composed of. Another example of a graph-based representation is the one proposed by Orio and Rodà \cite{orio2009}. In this approach, terminal nodes directly describe the music content, internal nodes represent its incremental generalisation, and arcs denote their relationships.

Vinet’s model can be further enriched with an additional category: vector representations. The main assumption underlying this category is that we need neither a symbolic representation nor a physical representation for describing music, but just vectors. This representation system is mainly employed for machine learning and deep learning applications.

Knowledge representations describe global characteristics of the music material, including objective descriptions such as the piece name, the music genre, the performing artists, the instruments played, as well as qualitative statements related to performance, sound quality, etc. \cite{vinet2004}.

Taken by its own, none of these MSRs seems to be capable of expressing the musical content in its completeness, but only in some specific and well-bounded aspects. Other expressive elements (e.g. timbre) remain the exclusive domain of signal representations.

The underlying idea of this research is to exploit Semantic Technologies as a means for connecting different types of MSR, obtaining a representation that can be structured and machine-readable but, at the same time, as expressive as possible.

\section{Related Work: Ontologies for Musicological Analysis}

Representing music by using the Semantic Web’s tools would allow interrelating and making interoperable different types of information that usually do not communicate between them. More in detail, this research’s main objective is to build an ontological model that can connect different music representation systems, namely symbolic representations with signal representations. Moreover, the model should be able to represent different types of music features that can assist musicological analysis.

Several ontologies specialised in the description of music and musicological content can be found on the web. The main works have been analysed to investigate the state-of-the-art of Semantic Web technologies for this type of application.

The Music Ontology \cite{raimond2007} is a formal framework for dealing with music-related information on the Semantic Web, including editorial, cultural, and acoustic information. This ontology is based on three other ontologies, namely Timeline Ontology\footnote{The Timeline Ontology (\url{http://motools.sourceforge.net/timeline/timeline.html})}, Event Ontology\footnote{The Event Ontology (\url{http://motools.sourceforge.net/event/event.html})}, and Functional Requirement for Bibliographic Records ontology (FRBR)\footnote{Functional Requirement for Bibliographic Records ontology (FRBR) (\url{https://www.ifla.org/publications/functional-requirements-for-bibliographic-records})}. The DOREMUS Ontology \cite{lisena2017}, is instead is an extension of FRBRoo\footnote{FRBRoo (\url{http://www.cidoc-crm.org/frbroo/home-0})} for describing cultural objects, applied to the specific domain of music. 

The MIDI Linked Data Cloud \cite{merono2017} proposes to use the data publishing method of Linked Data to interconnect symbolic music descriptions (to a great extent very similar to music scores) contained in MIDI files and is only suitable for describing music encoded in MIDI format. 
Similarly, the CHARM ontology \cite{harley2015} aims to describe abstract, multiple hierarchical musical structures, promoting interoperability between applications that perform search and analysis of musical material. This ontological model, however, is based on the CHARM specifications and only allows the description of music expressed in this specific symbolic representation.

The Music Theory Ontology (MTO) \cite{rashid2018} aims at including theoretical concepts that were not considered in previous music ontologies. This ontology focuses, indeed, on the rudiments that are necessary to understand and analyse music. This ontology can be considered expressive in representing musicological aspects. However, it is limited to the description of note attributes (dynamics, duration, articulation, etc.) at the level of detail of a note set. This peculiarity considerably reduces the expressiveness of the model, especially in relation to polyphonic and multi-voice pieces.

The Music Score Ontology (Music OWL) \cite{jones2017}, instead, aims to represent similar concepts described in the Music Theory Ontology. MusicOWL, however, focuses on music notation, and the classes it is composed of are all aimed at representing music in a notated form, hence related to the music sheet.

The Music Notation Ontology \cite{cherfi2017} is an ontology of music notation content, focused on the core ``semantic'' information present in a score and subject to the analytic process. Moreover, this work proposes to model some high-level analytic concepts (e.g. dissonances), and to associate them with fragments of the score. However, this research focuses only on symbolic music representations, not taking into account the audio signal. 

The WASABI project\footnote{Wasabi project (\url{http://wasabihome.i3s.unice.fr/})} \cite{meseguer-brocal2017}, instead, aims at the construction of a two million song knowledge base that combines metadata collected from music databases on the Web, metadata resulting from the analysis of song lyrics, metadata resulting from the audio analysis, and the development of semantic applications with high added value to exploit this semantic database. However, this study only relies on music data extracted from signal, which is prone to errors and uncertainties.

\section{The HaMSE Ontology}
\label{sec:hamse}

Starting from the lack of ontologies that deal with musicological research in an extensive way, we propose a novel ontological model that can be able to solve the issues listed in section 2.
This new model merges and harmonizes two or more different and non-interoperable music representations (e.g. audio data and scores), either symbolic or non-symbolic. This approach has multiple advantages: by connecting different MSRs, it will be possible to exploit all the characteristics that such representations can provide.
Secondly, this novel representation attempts to solve some of the problems that affect contemporary musicology. In particular, it should be able to infer new relationships that may allow a more inclusive approach between different disciplines that refer to musicology. Eventually, we envisage a formal approach to musicology, overcoming the strict dualism that very often occurs between computer science research and research in the humanities.

The ontology proposed will serve to formally encode music information using a standard interoperable representation, namely RDF/OWL knowledge graphs \cite{bizer2009}. 
Requirements for the ontology are also derived from algorithms and codes for data extraction.
In summary, it intends to address the following functional requirements:
\begin{romanenumerate}
\item representing music at different granularity levels
\item representing musicological features
\item connecting and harmonizing different music representations
\item connecting different representations with audio tracks of recordings of the same composition
\item linking a music piece to its perceptual, cultural, geographical and historical information.
\end{romanenumerate}

\subsection {Data Extraction}

Before moving to the ontology design phase, we have identified a series of features that could be useful for musicological analysis and for providing answers to the issues discussed in the previous sections. In order to be shaped appropriately by HaMSE, features need to be extracted from data.
The extraction can be divided into five main categories:
\begin{enumerate}
    \item Audio-to-score alignment
    \item Melodic feature extraction
    \item Harmonic feature extraction
    \item Structural feature extraction
    \item Emotional feature extraction
\end{enumerate}

Features are extracted from either the symbolic representation, or the audio track. The choice is made based on the characteristics of the representation systems and the quality of the results obtainable. In particular, harmonic and melodic features are extracted from symbolic representations, while emotional and structural features are extracted from signal representation. The audio-to-score alignment, instead, requires the use of both representations for its implementation. 
All codes are public and available on the GitHub repository of the project\footnote{The repository of the project is available at: \url{https://github.com/andreamust/HaMSE_Ontology}}. 

For linking a symbolic representation to a signal representation, we have aligned the two representations by means of a Dynamic Time Warping (DTW) algorithm. In particular, we have used the DTW algorithm provided by Librosa \cite{mcfee2015}, a Python package for audio and music signal processing. The symbolic representation is converted into signal by means of the midi2audio library, which converts files from MIDI format into wave files using the Fluidsynth synthesiser. Once converted, the algorithms align the two signals by comparing the chroma features extracted from them.

Melodic features have been extracted in the form of music patterns from the symbolic representations. A music pattern can be intended as a fragment that is repeated over the music piece. For this task, music21 was employed, an object-oriented toolkit for analysing, searching, and transforming music in symbolic (score-based) forms \cite{ariza2010}. This toolkit provides a modular approach, handling a great variety of notations and formats, such as MIDI, MuisicXML, HumDrum, etc. These representations, once imported into music21, are converted into Streams, nestable containers that allow researchers to find simultaneous events quickly, follow a voice, or represent instruments playing in different tempo and meters. 
The extracted patterns include:
\begin{itemize}
    \item Interval patterns
    \item Rhythmic patterns
    \item Melodic patterns
\end{itemize}
All these pattern types are extracted with or without rests.

Harmonic features are also extracted using music21. First of all, the chords are extracted from the set of harmonic features using music21’s Chordify function. With an approach similar to the one used for the extraction of melodic patterns, the chord progressions recurring within a song are extracted. The most frequent dissonances are also extracted, together with the key of the song.

Structural features are instead extracted from signal representation. To do this, a Librosa-based framework called MSAF \cite{nieto2016} is used. This library allows to extract a set of features and to analyse them by means of eight different boundary detection and segmentation algorithms.

Eventually, the emotional features are extracted employing a deep learning approach. A neural network is trained using the 4Q audio emotion dataset \cite{panda2018}, a dataset labelled on Russell’s Circumplex Model of Affect \cite{russell1980}, and predictions are then made on unseen data.

\subsection {Ontology Design}

The ontology development involves different phases to get a fully working and tested model. We have used eXtreme Design methods as described in \cite{blomqvist2010}, an agile approach to ontology engineering, based on scenarios, competency questions, design patterns, and unit tests. 
In order to determine the scope of the ontology, we defined a list of competency questions \cite{gruninger1995}.

The competency questions can be summarized as follows:
\begin{itemize}
    \item what is the tonality?
    \item who is the composer?
    \item what is the number of movements?
    \item how long (in seconds) is the song?
    \item how long (in bars) is the song?
    \item to which predominant category in Russell’s model of emotion does it belong?
    \item for a bar in a score, where is the correspondence in different audio recordings?
    \item how common is this pattern in songs that evoke a specific sentiment?
    \item is it a pattern most common in a specific predominant category in Russell’s model of emotion?
    \item for a predominant category in Russell’s model of emotion, what are the most common patterns that belong to it?
    \item which are the most common structures in songs that contain a specific pattern?
    \item which composer wrote the greatest number of songs containing a given pattern?
\end{itemize}

The ontology specifically created for this research is called HaMSE (Harmonic, Melodic, Structural and Emotional features ontology)\footnote{The ontology documentation is available at: \url{https://andreapoltronieri.org/HaMSE_project/}}, and its rdf/owl file is available at the following URI: \url{https://purl.org/andreapoltronieri/HaMSEontology}. 

We created a small knowledge base in order to test the ontology. The knowledge base was populated with data from a Bach's chorale, namely the sixth movement of the church cantata \emph{``Wie schön leuchtet der Morgenstern''}. The piece is catalogued under the index BWV1. The sixth and last movement is the chorale entitled \emph{``Wie bin ich doch so herzlich froh''}. 
The ontology was then tested against the competency questions, running SPARQL queries onto the created knowledge base. 

The ontology has the \texttt{hamse:MusicologicalFeature} class as a top class. This is because the ontology schema is aimed at representing the musicological features as the main elements of the analysis. From this top class, only the classes related to the audio-score alignment are excluded. The information related to the musician and the composition are also considered musicological features, as their inclusion is conditional on their use, i.e. musicological analysis.

The modelling of the ontology takes into account both the music as represented on the score and as computer-readable symbolic representation. Furthermore, each of the events (i.e. notes and rests) expressed in the symbolic notation is linked to the signal representation, which can provide more precise information on timbre and dynamics. 

For clarity purpose, the ontology is here divided into small blocks, defined according to the entities represented. 

\paragraph*{Describing the Musical Work}

First of all, the ontology describes the editorial and cutural information of the music piece. To do this, we decided to reuse the Music Ontology. This choice is due to its wide adoption, and its interoperability with other ontologies (e.g. TimeLine ontology, Music Score Ontology, etc.).
The part of the Music Ontology that we have reused focuses on the concept of music work (\texttt{mo:MusicalWork}), which is a subclass of \texttt{frbr:Work}. 

The Musical Work is linked to a composition event (\texttt{mo:Composition}) which in turn is linked to the composer, expressed through the \texttt{frbr:Agent} class. The composition can generate a score (\texttt{mo:Score}), which may in turn have been published, and which may therefore be expressed by the class \texttt{mo:PublishedScore}. The Musical Work also has a number of relationships to a specific arrangement \\(\texttt{mo:Arrangement}), to the musical genre that defines it (\texttt{mo:Genre}), and to the dominant key of the composition (\texttt{mo:Key}). Furthermore, this Musical Work is linked to the performance of the work itself (\texttt{mo:Performance}), which is defined in relation to the instruments used (\texttt{mo:Instrument}), the musicians and the composer.

Finally, the performance is connected to the produced sound (\texttt{mo:Sound}), which in turn is connected to the generated recording (\texttt{mo:Recording}). The latter produces a signal (\texttt{mo:Signal}), which can be either digital (\texttt{mo:DigitalSignal}) or analogue (\texttt{mo:AnalogSignal}). The signals are also interconnected, since the digital signal is the product of the sampling of the analogue signal.

\paragraph*{Describing Symbolical Music Events}

\begin{figure}[!ht]
    \centering
    \includegraphics[width=\textwidth]{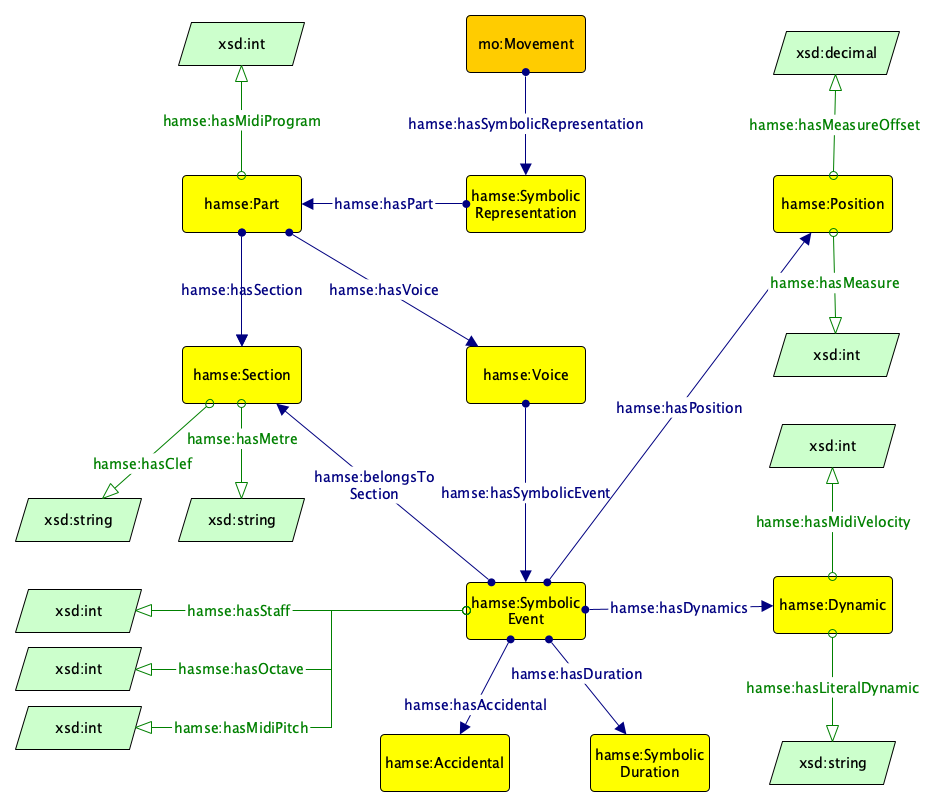}
    \caption{Detail of the ontology part representing the symbolic music events}
    \label{fig:Representing symbolic events}
\end{figure}

In order to model the symbolic representations of a composition, the HaMSE ontology is proposed (see Figure \ref{fig:Representing symbolic events}). 

With the class \texttt{hamse:SymbolicEvent}, we describe everything that can be a nuclear element of the symbolic representation, intended as either a note or a pause. The notation event, such as all the other classes listed in this section, is intended as a subclass of \texttt{hamse:MusicologicalFeature}. 

Each characterizing element of the note is expressed with the corresponding MIDI feature value. The presence of these MIDI indications allows first of all to have a higher level of detail, but also to allow this representation to be expanded and aligned with other representations in the future. Since MIDI is a standard in computer music domain, these references are also useful for having precise indications regarding elements such as pitch and tempo, as well as the instruments used and metre information.

This part of the ontology is connected to the Music Ontology by means of the class \texttt{mo:Movement}: for each movement, in fact, it is possible to associate different symbolic representations which are described by the class \texttt{hamse:SymbolicRepresentation}. Each representation may contain several parts (\texttt{hamse:Part}) which are associated with the MIDI instrument number via the datatype property \texttt{hamse:hasMidiProgram}. Each part is then fragmented into its different voices (\texttt{hamse:Voice}) and sections (\texttt{hamse:Section}). A section is a grouping of notes that share the same metre and clef, which are described by the datatype properties \texttt{hamse:hasMetre} and \texttt{hamse:hasClef}, respectively. Each voice and part then contain the music event, expressed by the class \texttt{hamse:SymbolicEvent}. This class has eight different subclasses, one for each natural note plus a class for rests.

The position, pitch, note accidentals, dynamics and duration are described in detail for each Symbolic Event. Accidentals are expressed by the class \texttt{hamse:Accidental}. The note position within the piece is expressed with a complex set of relations that describe both the abstract position of the note and the position of the note in seconds. 
The Position class is then linked to the TimeLine ontology by means of the class \texttt{tl:Interval}, which links the duration of the note to its position on a (abstract) timeline. 

The class \texttt{hamse:Dynamic} expresses the dynamic information of the note, by linking the Symbolic Event to the textual dynamic values, through the datatype property \\\texttt{hamse:hasLiteralDynamic}, and to the numerical values related to MIDI velocity, via the datatype property \texttt{hamse:hasMidiVelocity}.

The note's pitch is also expressed in two forms: by means of the MIDI number of the note (datatype property \texttt{hamse:hasMidiPitch}), and by the references to the score the note belongs to (datatype properties \texttt{hamse:hasOctave} and \texttt{hamse:hasStaff} respectively).

\paragraph*{Representing Extracted Features}

\begin{figure}[!ht]
    \centering
    \includegraphics[width=\textwidth]{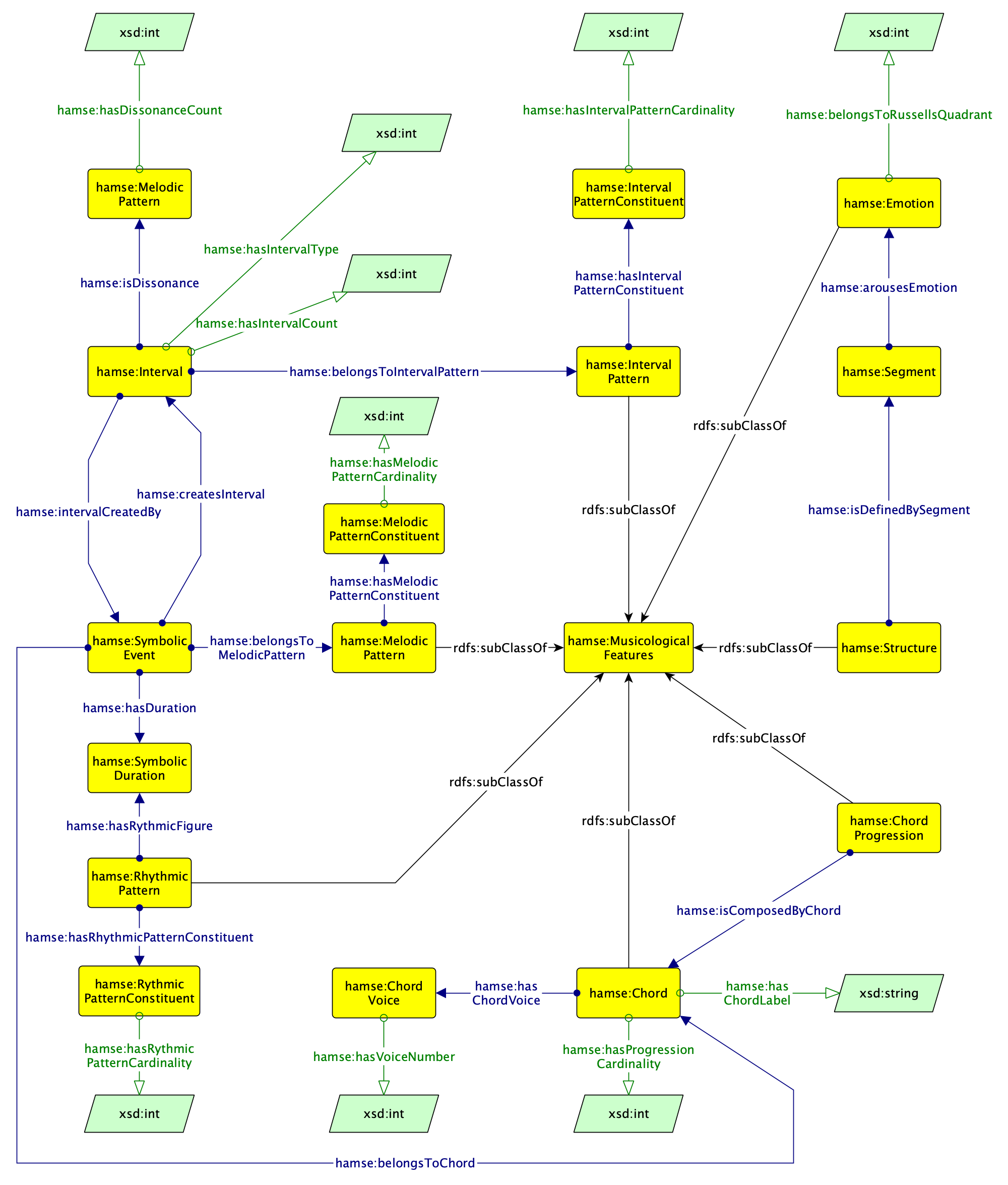}
    \caption{Detail of the ontology part representing the extracted features}
    \label{fig:Extracted features}
\end{figure}

The representation of the extracted features (see Figure \ref{fig:Extracted features}) aims to connect these features with both the symbolic representation and the signal representation, providing a temporal reference of their location. 

The class \texttt{hamse:MusicologicalFeatures} has as subclasses all the features that have been extracted from both the audio and the symbolic representation. 

\newpage
\noindent{The Feature extracted are described by the classes:}
\begin{itemize}
    \item\texttt{hamse:MelodicPattern} \item\texttt{hamse:Emotion}
    \item\texttt{hamse:IntervalPattern} \item\texttt{hamse:Chord} \item\texttt{hamse:RhythmicPattern} \item\texttt{hamse:ChordProgression}  \item\texttt{hamse:Structure}. 
\end{itemize}

All these classes, as subclasses of \texttt{MusicologicalFeature}, are linked to the \\\texttt{tl:AbstractInterval} class, for describing the feature position within the score, and find the correspondent time interval on the audio signal timeline.

\paragraph*{Score-audio-features Alignment}

The TimeLine Ontology has been used to align the symbolic representation to the audio track. In particular, in this case we aligned an abstract timeline (\texttt{tl:AbstractTimeLine}) with a discrete timeline (\texttt{tl:DiscreteTimeLine}). 

These two different timelines are aligned by means of \texttt{tl:TimeLineMap}. The intervals in which events (both notes and features) occur are defined from the offset second to the end second and are connected to the timeline, either abstract or discrete.

\subsection{Conclusions and Future Developments}

We have proposed a novel ontology to support musicological analysis. Our contribution addresses: the harmonization of multiple music representations, fostering interoperability, and the representation of musicological features.
To do this, a set of features have been identified, and algorithms to extract them have been presented. These different features can capture musical elements that aim to bridge different music research fields. They can also serve as a link for research in other fields related to music, first of all, psychoacoustics.

Furthermore, this study highlights the complexity of musical and musicological analysis. In particular, the manifold nature of musical content has been emphasised, as well as how Semantic Web technologies can be an excellent tool for representing different aspect of the musical material and making them interoperable.

A future development of this work may be the population of the ontology with a large number of data leading to the creation of a knowledge graph. This will involve the implementation of algorithms that will automatically populate the KG, converting the extracted features into RDF. 

A further improvement would be the modularisation of the ontology, whose size, usage, and maintenance are ideally controlled by means of a network of modules. 

It is also possible to evolve the ontology, and adapt it to different styles of composition and notation. At present, in fact, the ontology can only describe pieces that are part of the Western tonal tradition. The inclusion of classes and properties (e.g. related to microtonal pitch variations) could expand the ontology model's scope.

HaMSE ontology is based on musicological features, which is a concept functional to musicological analysis, hence focused on symbolic notation. It is possible to extend the ontology to represent sound as an entity. This would allow to treat knowledge about sound as a realisation of the symbolic representation by a musician or a machine.


\bibliography{references.bib}

\end{document}